\documentclass[]{revtex4}
 \usepackage[]{natbib}
\usepackage{graphicx}
\usepackage{amsmath,amssymb}
\usepackage{longtable}
\usepackage{booktabs}
\usepackage{multirow}
\usepackage{fancyvrb}
\usepackage[utf8]{inputenc}
\begin{document}
\title[FES for prediction of PC]{Fuzzy expert system for prediction of prostate cancer}

\author{Juthika Mahanta$^{1,}$\footnote{juthika@math.nits.ac.in}}
\author{Subhasis Panda$^{2,}$\footnote{subhasis@phy.nits.ac.in}}

\affiliation{$^1$Department of Mathematics,
              National Institute of Technology Silchar, Silchar, Cachar, Assam, India, Pin-788010}
\affiliation{$^2$Department of Physics,
              National Institute of Technology Silchar, Silchar, Cachar, Assam, India, Pin-788010}
  
\begin{abstract}
A fuzzy expert system (FES) for the prediction of prostate cancer (PC) is prescribed in this article. Age, prostate-specific antigen (PSA), prostate volume (PV) and $\%$ Free PSA ($\%$FPSA) are fed as inputs into the FES and prostate cancer risk (PCR) is obtained as the output.  Using knowledge based rules in Mamdani type inference method the output is calculated. If PCR $\ge 50\%$, then the patient shall be advised to go for a biopsy test for confirmation. The efficacy of the designed FES is tested against a clinical data set. The true prediction for all the patients turns out to be $68.91\%$ whereas only for positive biopsy cases it rises to $73.77\%$. This simple yet effective FES can be used as supportive tool for decision making in medical diagnosis.
\end{abstract}

\maketitle

\section{Introduction}\label{intro}
Artificial intelligence (AI) is defined as the intelligence processed by machines. With the advancement in the computer system, machines exhibit tasks which normally need human intelligence. Development of AI techniques has revolutionized many areas like robotics, transportation, education, marketing etc. including medical diagnosis and health care. Medical diagnosis deals with the analysis of complex medical data. The primary job in medical diagnosis is to reach to a decision using expert's logical reasoning. Handling large complex data and many uncertainties make this job very difficult. AI appears to be very handy for this job. AI in medical diagnosis has added expert human reasoning in simulation of computer-aided diagnosis process. There are different AI methods used in medical diagnosis, fuzzy logic is one of the most popular one. AI is the technique of mimicking human intelligence by the help of advance computer systems. Human brain takes natural languages as inputs which aren't feasible to be represented by Boolean logic (either true or false). So, there is a requirement of representation outside these two possibilities. Fuzzy logic exactly does that. Fuzzy logic appears closer to the way human brain works. Therefore, in AI, fuzzy logic shows the sign of being a natural choice. \\
Uncertainties and imprecision are connected with every aspect of our day to day life activities. Specifically in medical diagnosis domain, one encounters a lot of uncertainty and vagueness. It becomes very difficult to identify a particular disease from the said symptoms of the patients, as it contains lot of approximate and inaccurate information. On the other hand, a particular symptom can possibly lead to many different disjoint diseases, whereas for the same disease, the symptom may manifest itself in completely different ways from person to person. There are inherent uncertainties in the process of decision making in medical diagnosis, even for an expert too. To tackle these inexact, linguistic inputs, fuzzy logic based expert system is turned out to be very useful. The concepts of fuzzy set and fuzzy logic were introduced by Prof. L. A. Zadeh \cite{zadeh}. In contrast to binary logic, fuzzy logic deals with multi-valued logic which is a mathematical tool to represent the real world effectively. Due to its usefulness and simplicity, fuzzy logic has drawn huge attention of interdisciplinary researchers round the globe.\\
Fuzzy logic based expert systems are widely used in many areas of medical diagnosis and decision-making process. Particularly, in the area of prostate cancer, very few literature are available \cite{saritas,benecchi,yuksel,seker,kar,castanho,abbod_review,lorenz}. These literature address the problem in different angles and also use fuzzy logic in disjoint ways. Some researchers have used hybrid system to treat this problem. We focus our attention to fuzzy logic based expert system to predict the prostate cancer risk. From careful study of the literature, we found that $\%$FPSA is a very crucial parameter, along with age, PSA and PV for early detection of PC. Therefore, we formulate a fuzzy expert system by taking care of all these inputs. \\   
The paper is organized as follows. Section \ref{fes} describes the fuzzy expert system, where all the inputs and the output are discussed. In section \ref{res}, we apply our FES to a medical data set and discuss our findings. Finally we summarize and conclude in section \ref{con}.
\section{Fuzzy Expert System (FES)} \label{fes}
\begin{figure}[h!]
\includegraphics[width=\textwidth]{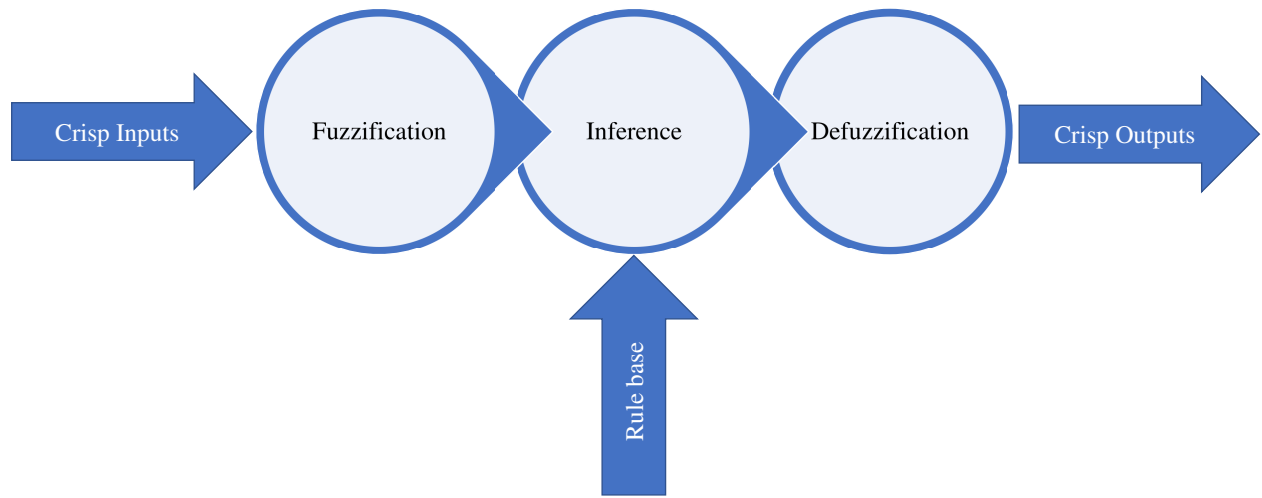}
\caption{General architecture of a FES.}
\label{fig:0}      
\end{figure}

An expert system which uses fuzzy logic instead of Boolean logic is called fuzzy expert system (FES). A FES is a form of artificial intelligence which deals with membership functions and some prescribed rule base to evaluate a set of data. Fuzziness is introduced to the crisp inputs of a FES by means of suitable membership functions. Once membership functions are defined for all input variables, then they are fed to a particular inference method for further action. Here, we have used Mamdani (max-min) inference method which is most popular in literature. Rules of the FES developed here are of IF-THEN form. Mamdani type inference method results in fuzzy sets as output. For a given set of input values, some relevant rules will be fired to produce a fuzzy output in Mamdani type inference method. Fuzzy output is defuzzified using different techniques to obtain crisp output. Centroid method is used for defuzzification in our FES. General structure of a FES is shown in below figure \ref{fig:0}.

\noindent We have used medical data of the patients as given in reference \cite{saritas}. Depending on the data set, the range of different inputs are determined. In the following two subsections, we discuss about different inputs and the output of the FES.
\subsection{Input variables}
\subsubsection{Age}
\begin{figure}[h!]
\includegraphics[width=\textwidth]{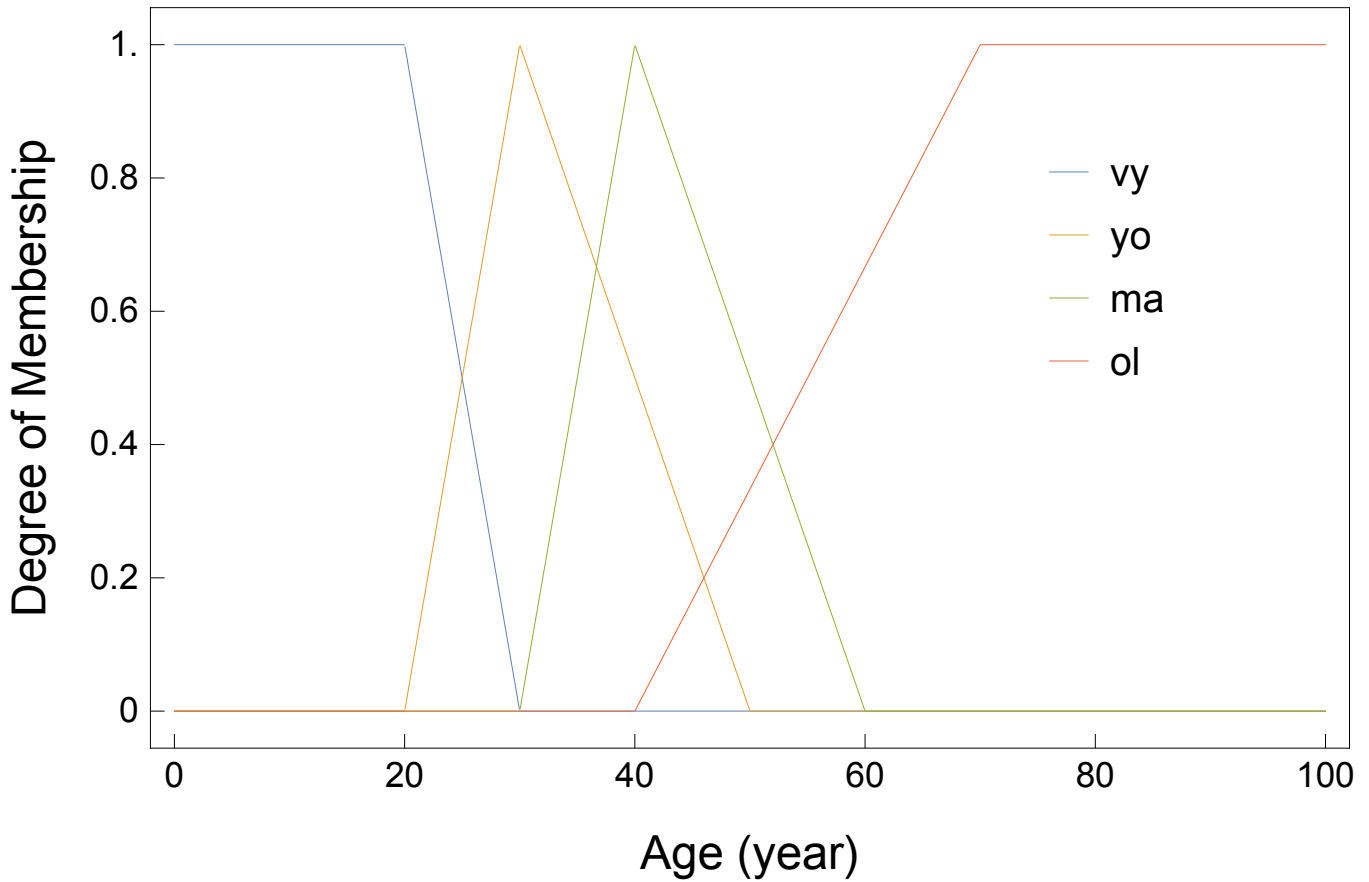}
\caption{Membership functions for ``age".}
\label{fig:1}      
\end{figure}

Age of man is an important parameter to calculate the risk factor for prostate cancer. For a man having no family history of PC, the chance of getting it increases after the age of 50. This number changes from race to race. However, two out of three PCs are diagnosed in men at the age of 65 or above. The input variable ``age" is represented by four fuzzy sets, namely, ``very young", ``young", ``middle age" and ``old". First and fourth fuzzy sets are represented by trapezoidal membership functions whereas for second and third we have used triangular membership functions. Table \ref{tab:1} lists crisp sets and the corresponding fuzzy sets for the input ``age". The membership functions for the same are plotted in figure \ref{fig:1}.
\begin{table}[h!]
\caption{Fuzzification of the input variable ``age".}
\label{tab:1}
\begin{tabular}{lll}
\hline\noalign{\smallskip}
Input variable & Crisp set & Fuzzy set  \\
\noalign{\smallskip}\hline\noalign{\smallskip}
Age (year) & 0-30 & Very young \\
 & 20-50 & Young \\
 & 30-60 & Middle age \\
 & 40-100 & Old \\
\noalign{\smallskip}\hline
\end{tabular}
\end{table}
\subsubsection{Prostate-specific Antigen (PSA)}
\begin{figure}[h!]
\includegraphics[width=\textwidth]{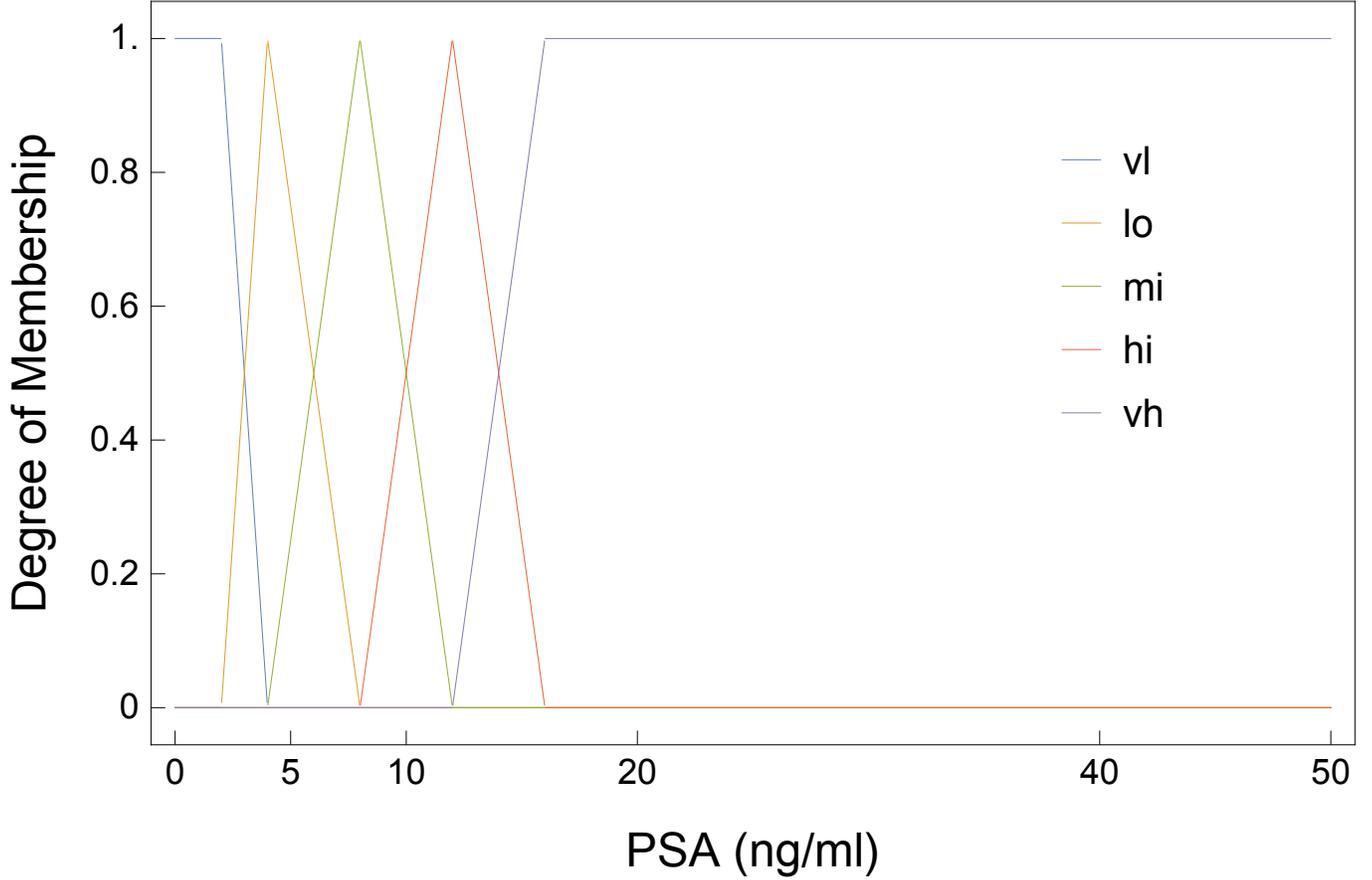}
\caption{Membership functions for ``PSA".}
\label{fig:2}      
\end{figure}

PSA has altered drastically the management of prostate cancer in men. The PSA test for blood can provide early stage detection of PC \cite{brawer99}. PSA is a protein secreted by the prostate gland which helps to keep the semen in liquid form. Some parts of this protein will pass into blood which give rise to the increase in normal PSA level. Elevation in PSA level in blood depends up on the health of prostate gland and the age of the person. A healthy prostate will release less PSA in blood compared to a cancerous gland. So, a rise in PSA level over normal range could be a possible indicator of PC. Although, elevated PSA level may be caused due to other factors like acute bacterial prostatitis, enlargement of prostate and other urinary retention. The measurement of PSA is expressed as nanograms per milliliter of blood. The normal range of PSA number can be age specific and also race specific. The input variable ``PSA" is represented by five fuzzy sets, e.g., ``very low", ``low", ``middle", ``high" and ``very high". For the first and fifth sets we have used trapezoidal membership functions while for the rest, triangular membership functions are used. In table \ref{tab:2}, we have shown the crisp sets and the corresponding fuzzy sets. The plot of the membership functions for the input ``PSA" is displayed in figure \ref{fig:2}.  
\begin{table}[h!]
\caption{Fuzzification of the input variable ``PSA".}
\label{tab:2}
\begin{tabular}{lll}
\hline\noalign{\smallskip}
Input variable & Crisp set & Fuzzy set  \\
\noalign{\smallskip}\hline\noalign{\smallskip}
PSA (ng/ml) & 0-4 & Very low \\
 & 2-8 & Low \\
 & 4-12 & Middle \\
 & 8-16 & High \\
 & 12-50 & Very high\\
\noalign{\smallskip}\hline
\end{tabular}
\end{table}
\subsubsection{Prostate Volume (PV)}
\begin{figure}[h!]
\includegraphics[width=\textwidth]{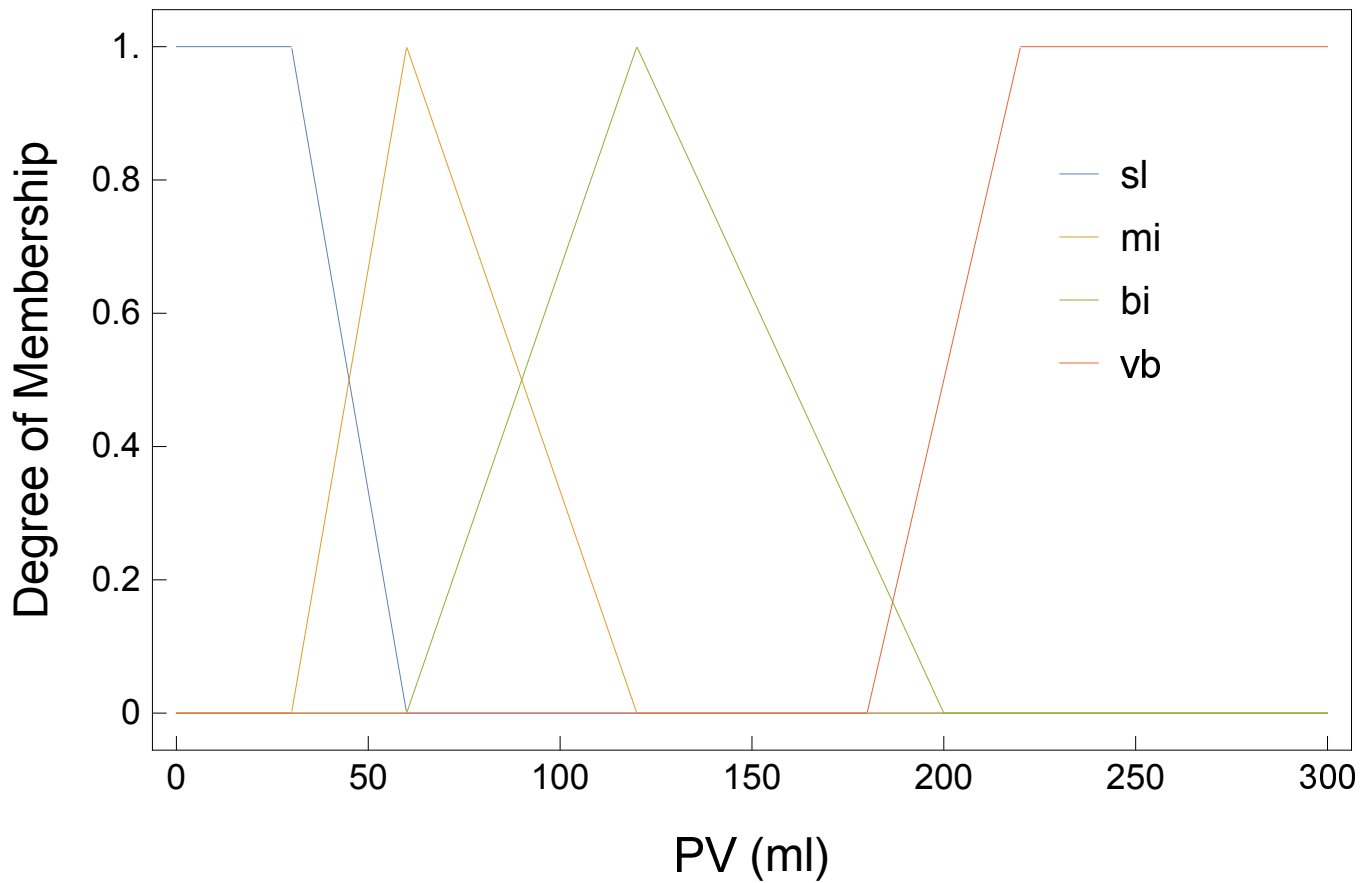}
\caption{Membership functions for ``PV".}
\label{fig:3}      
\end{figure}

A healthy human male's prostate is marginally larger than a walnut. It is a crucial parameter for early detection of PC. There is a characteristic pattern in the growth of prostate with age. That pattern can change from race to race. With the increase in the prostate volume there is a possibility of sampling error in systematic sextant needle biopsy. It is wise to use prostate volume as a factor while determining the necessity to repeat biopsy with initial negative result \cite{brawer_pv}. Prostate is mainly divided in four zones in pathological terminology and total prostate volume as well as transition zone volume are measured in ultrasound. According to transrectal ultrasound (TRUS) guidance \cite{zhang}, prostate width ($W$) (maximal transverse diameter) is estimated on an axial image. Prostate length ($L$) (longitudinal diameter, the distance between proximal external sphincter and urinary bladder) and height ($H$) (maximal antero-posterior diameter) are measured on a mid-sagittal image \cite{bangma}. The total prostate volume (TPV) is calculated using the prolate elliptical formula,  TPV = $\frac{\pi} {6} \times W \times L \times H$. The transition zone volume can be calculated using the same formula by measuring the required dimensions from the ultrasound. The length, width and height changes with the age. The rate of change for the length is significant compared to other two dimensions as the man approaches the age 60. So, based on the above information we have divided the input ``prostate volume" in four fuzzy sets such as ``small", ``middle", ``big" and ``very big". Trapezoidal and triangular membership functions are used to represent them. Crisp sets, fuzzy sets and the membership functions are listed in table \ref{tab:3} and figure \ref{fig:3} respectively for the input variable ``prostate volume".
\begin{table}[h!]
\caption{Fuzzification of the input variable ``PV".}
\label{tab:3}
\begin{tabular}{lll}
\hline\noalign{\smallskip}
Input variable & Crisp set & Fuzzy set  \\
\noalign{\smallskip}\hline\noalign{\smallskip}
PV (ml) & 0-60 & Small \\
 & 30-120 & Middle \\
 & 60-200 & Big \\
 & 180-300 & Very big \\
\noalign{\smallskip}\hline
\end{tabular}
\end{table}
\subsubsection{Percentage of Free PSA ($\%$FPSA)}
\begin{figure}[h!]
\includegraphics[width=\textwidth]{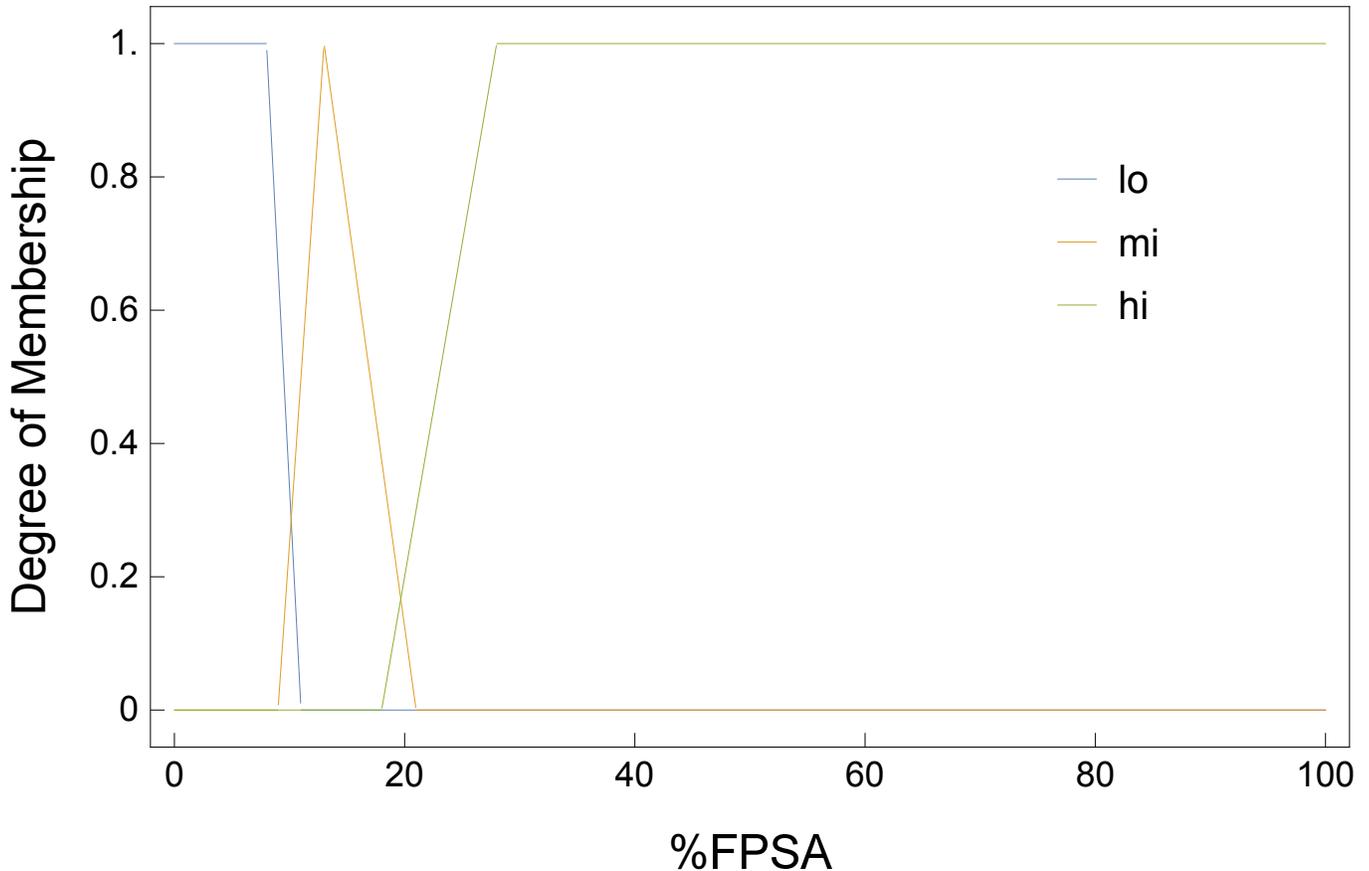}
\caption{Membership functions for ``$\%$FPSA".}
\label{fig:4}      
\end{figure}

PSA is a protein which exists in different forms in serum. PSA circulates through body via bound to some other proteins or in unbound form. Free PSA test measure the ratio of the unbound PSA to bound PSA whereas normal PSA test measure the total PSA (both bound and unbound) level in blood \cite{labmed,ito}. $\%$FPSA is calculated as $\frac{{\rm Free ~ PSA}}{{\rm Total PSA}} \times 100\%$. As we already discussed that PSA level may rise not only due to cancerous prostate but because of many other reasons. Therefore, for the early detection of the PC, which is potentially in its curable stage, requires a lower cutoff for PSA level, gives rise to avoidable biopsies. It is also observed that the men with PC are likely to have $\%$FPSA lower than those of benign disease \cite{labmed,catalona}. So, along with an elevated PSA level, $\%$FPSA cutoff will be a good indicator for the early stage detection. Based on this we have categorized the input variable ``$\%$FPSA" in three fuzzy sets, namely, ``low", ``middle" and ``high". For their representation we have used trapezoidal and triangular membership functions. Crisp sets, fuzzy sets and the membership functions are shown in table \ref{tab:4} and figure \ref{fig:4}.  
\begin{table}[h!]
\caption{Fuzzification of the input variable ``$\%$FPSA".}
\label{tab:4}
\begin{tabular}{lll}
\hline\noalign{\smallskip}
Input variable & Crisp set & Fuzzy set  \\
\noalign{\smallskip}\hline\noalign{\smallskip}
$\%$FPSA & 0-11 & Low \\
  & 9-21 & Middle \\
 & 18-100 & High \\
 \noalign{\smallskip}\hline
\end{tabular}
\end{table}
\subsection{Output Variable}
\subsubsection{Prostate Cancer Risk (PCR)}
\begin{figure}[h!]
\includegraphics[width=\textwidth]{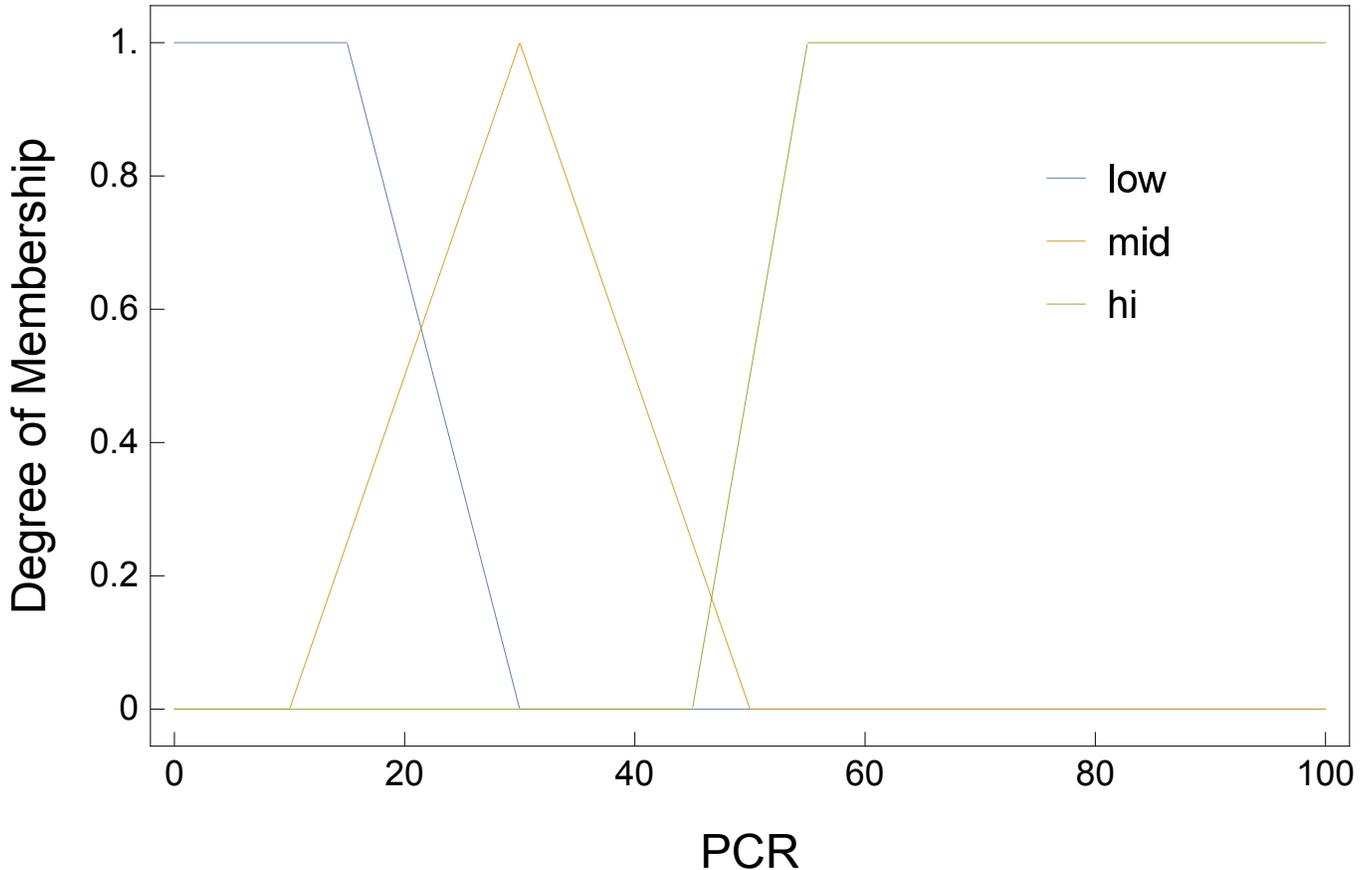}
\caption{Membership functions for ``PCR".}
\label{fig:5}      
\end{figure}

The output of the FES is PCR. Numerical value of it will help us to identify the benignant one or the malignant one by considering PSA, age, PV and $\%$FPSA of the patient as input variables for the FES. If the PCR $\%$ value is $\ge 50$, then we can anticipate that the patient has a high chance of having PC. Therefore, he will be recommended for biopsy test for confirmation. The positive biopsy result confirms our prediction whereas negative result contradicts. The case of PCR $\%$ value $<50$ can be understood in similar fashion. The output variable, PCR is categorized into three fuzzy sets, namely, ``low", ``middle" and ``high". Trapezoidal and triangular membership functions are used to represent them. The table \ref{tab:5} and the figure \ref{fig:5} are displaying these fuzzy sets and their membership functions respectively.
\begin{table}[h!]
\caption{Fuzzification of the output variable PCR.}
\label{tab:5}
\begin{tabular}{lll}
\hline\noalign{\smallskip}
Output variable & Crisp set & Fuzzy set  \\
\noalign{\smallskip}\hline\noalign{\smallskip}
Prostate Cancer Risk ($\%$) & 0-30 & Low \\
 & 10-50 & Middle \\
 & 45-100 & High \\
\noalign{\smallskip}\hline
\end{tabular}
\end{table}
\subsection{Fuzzy Rule Base}
\begin{figure}[h!]
\includegraphics[width=0.63\textwidth]{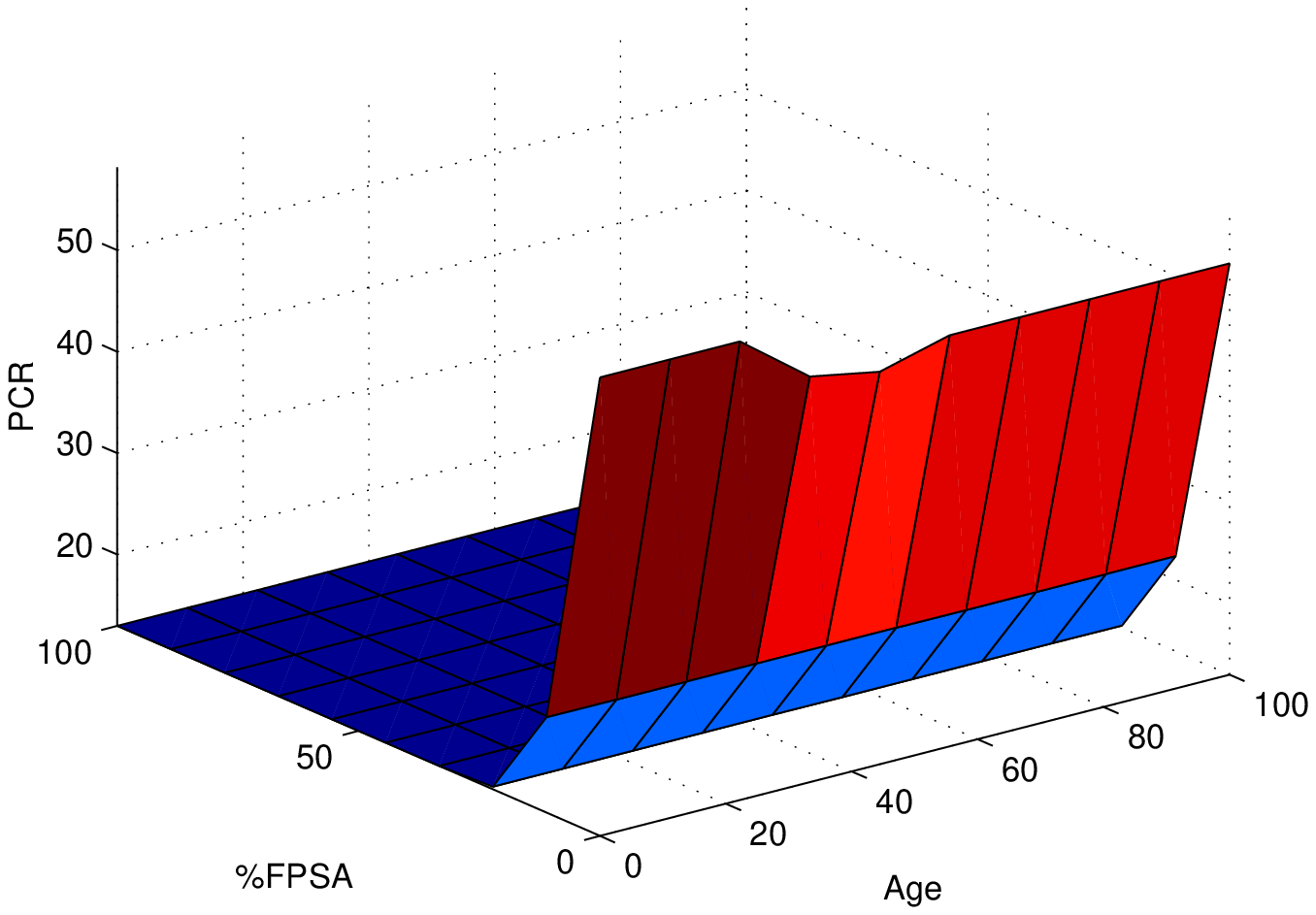}
\includegraphics[width=0.6\textwidth]{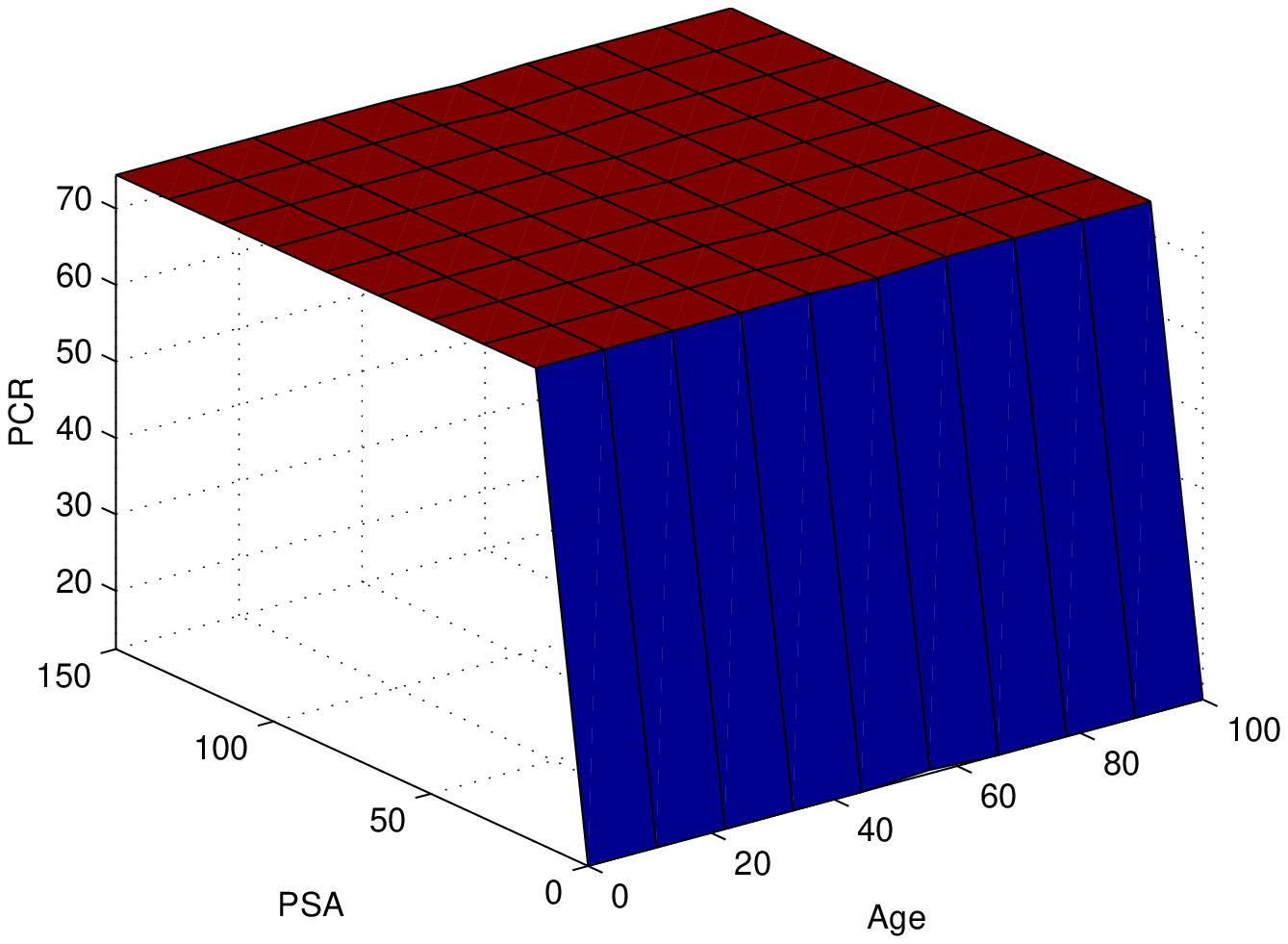}
\caption{Surface plot of four-input FES  (left) by taking two variables as inputs and PCR as output while keeping reference values for other two inputs. For the figure on the left, age \& $\%$FPSA are inputs and reference values for PV \& PSA are $100$ and $6$ respectively. The figure on the right, we have taken PSA \& age as inputs while fixing reference values for PV and $\%$FPSA at 100 and $15$ respectively. }
\label{fig:7}      
\end{figure}

In this FES, four input variables are age, PSA, PV and $\%$FPSA which are represented by four, five, four and three membership functions respectively. So, our FES has total $4\times5\times4\times3=240$ rules to estimate PCR as the output, which is also characterized by three membership functions. Some of the selected rules are displayed below.
\begin{Verbatim}[fontsize=\footnotesize]
1. If (Age is vy) and (PSA is vl) and (PV is small) and (%FPSA is low) then (PCR is low) (1)     
8. If (Age is vy) and (PSA is vl) and (PV is big) and (%FPSA is mid) then (PCR is low) (1)       
65. If (Age is yo) and (PSA is vl) and (PV is middle) and (%FPSA is mid) then (PCR is low) (1)   
97. If (Age is yo) and (PSA is hi) and (PV is small) and (%FPSA is low) then (PCR is high) (1)   
130. If (Age is ma) and (PSA is vl) and (PV is verybig) and (%FPSA is low) then (PCR is low) (1) 
172. If (Age is ma) and (PSA is vh) and (PV is middle) and (%FPSA is low) then (PCR is high) (1) 
196. If (Age is ol) and (PSA is lo) and (PV is middle) and (%FPSA is low) then (PCR is low) (1)  
240. If (Age is ol) and (PSA is vh) and (PV is verybig) and (%FPSA is high) then (PCR is mid) (1)
\end{Verbatim}
For example, rule $1$ can be elucidated as, if age of the patient is very young, PSA is very low, PV is small and $\%$FPSA is low then PCR for that patient is low. The wieght of this rule is maximum which is specified by number $1$ in the parentheses after the rule. Other rules can be spelled out in similar manner. Surface plot gives a visualization of the rules applied in FES. Since, our FES has four inputs and one output, it is not possible to plot the dependence of the output against all inputs. So, in surface plot we have plotted the output against any two input variables by keeping some reference values for the rest two inputs. Fig \ref{fig:7} displays two such surface plots where relevant parameters are mentioned in the caption. For example, PCR is high (very high) for $\%$FPSA value less than $20\%$ ($10\%$) irrespective of the patient's age. 

\subsection{Defuzzification \& Mamdani Inference Engine}
\begin{figure}[h!]
\includegraphics[width=1.2\textwidth]{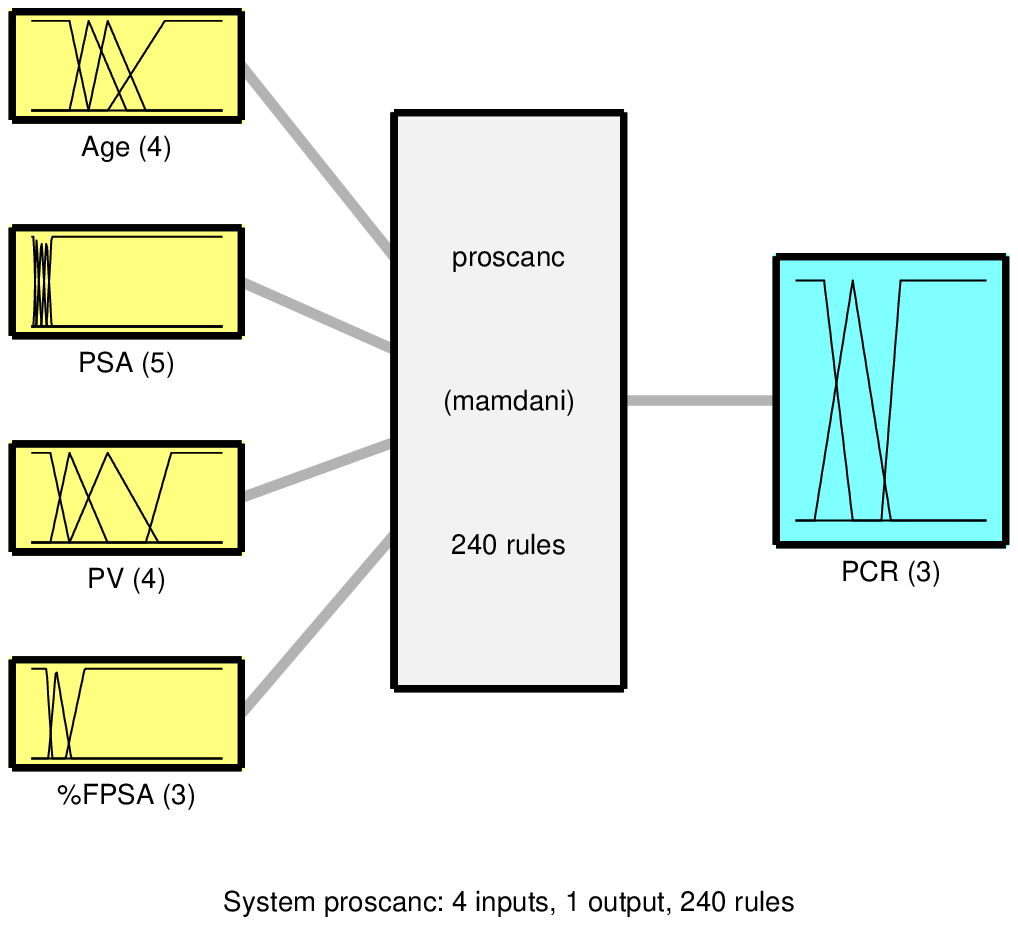}
\caption{FES structure}
\label{fig:6}      
\end{figure}

Defuzzification is the process of obtaining a precise quantity from a fuzzy set. In this FES, we have employed centroid method to defuzzify which is given by 
\begin{equation}
{\bar z}=\frac{\int \mu_C(z)\cdot z~ dz}{\int \mu_C(z)~dz}. 
\end{equation}
For a given set of input variables, viz. age=$56$, PSA=$9.05$, PV=$39$ and $\%$FPSA=$8.51$, we calculate the corresponding degrees of membership by identifying the proper fuzzy sets for each input variable. This particular choice of inputs will fire, say $k$ number of rules. Truth degree ($\alpha_i$) of the $i^{\rm th}$ firing rule is determined by taking min of corresponding membership values of each input variable. Thereafter, max of all $\alpha_i$ will be the membership value of PCR. This is how Mamdani max-min inference technique is used. Then applying the centroid method, we obtain crisp value of PCR. Steps are shown below for the above mentioned input data.
\begin{enumerate}
    \item Age = $56$, $\mu_{\rm ma}(56)=0.2$ and $\mu_{\rm ol}(56)=0.533$.
    \item PSA = $9.05$, $\mu_{\rm mi}(9.05)=0.7375$ and $\mu_{\rm hi}(9.05)=0.2625$.
    \item PV = $39$, $\mu_{\rm sl}(39)=0.7$ and $\mu_{\rm mi}(39)=0.3$.
    \item $\%$FPSA = $8.51$, $\mu_{\rm lo}(8.51)=0.83$.
\end{enumerate}
For the above set of input data, eight rules will come into action and yield
\begin{enumerate}
\item $\alpha_{145}={\rm min}(0.2,0.7375,0.7,0.83)=0.2$,
\item $\alpha_{148}={\rm min}(0.2,0.7375,0.3,0.83)=0.2$,
\item $\alpha_{157}={\rm min}(0.2,0.2625,0.7,0.83)=0.2$,
\item $\alpha_{160}={\rm min}(0.2,0.2625,0.3,0.83)=0.2$,
\item $\alpha_{205}={\rm min}(0.533,0.7375,0.7,0.83)=0.533$,
\item $\alpha_{208}={\rm min}(0.533,0.7375,0.3,0.83)=0.3$,
\item $\alpha_{217}={\rm min}(0.533,0.2625,0.7,0.83)=0.2625$,
\item $\alpha_{220}={\rm min}(0.533,0.2625,0.3,0.83)=0.2625$,
\end{enumerate}
which further provides
\begin{flalign*}
  \alpha &=  {\rm max}(\alpha_{145},\alpha_{148},\alpha_{157},\alpha_{160},\alpha_{205},\alpha_{208},\alpha_{217},\alpha_{220}), \\
  &=  {\rm max}(0.2,0.2,0.2,0.2,0.533,0.3,0.2625,0.2625)= 0.533. 
\end{flalign*}
Using $\alpha = 0.533$, and applying centroid method we get the PCR is $74.01\%$, which is much greater than our cutoff value of $50\%$. Therefore, in this case, the patient should be advised to go for a biopsy to confirm.
We have used fuzzy logic toolbox of Matlab software for calculation of the entire set of medical data of 119 patients. Block diagram of FES is portrayed in the figure \ref{fig:6}.
\section{Results and Discussion} \label{res}
We have applied our proposed FES to analyze the data (given in reference \cite{saritas}) presented in table \ref{tab:6}. Here, the range of different inputs are chosen depending upon the minimum and the maximum values of the respective input variables. For example, for the input ``PV", the maximum value is $235$ ml and the minimum value is $15$ ml. Therefore, we have chosen the range for ``PV" to be $0-300$ ml. Feeding the four inputs into the FES, we obtain PCR as the output. If the value of PCR is greater or equal to $50\%$, then the patient should be advised to go for a biopsy test to confirm whether the prostate problem is of benignant or malignant type. We have compared outcomes of our FES with the results of biopsy. Out of total $119$ patients, $61$ patients had positive biopsy results, while the rest $58$ patients had negative results. The true prediction by our FES is $68.91\%$. The proposed FES has correctly predicted the positive biopsy results for $45$ patients ($73.77\%$) and the negative biopsy results for $37$ patients ($63.79\%$). True prediction percentage of our FES is better than that of the existing systems for prediction in literature. Saritas {\it et. al} \cite{saritas} has claimed $64.71\%$ true prediction for the same set of data, while online calculator for PC prediction and FPSA/PSA ratio provide only $62.18\%$ and $60.50\%$ respectively. Inclusion of one parameter has definitely increased the number of rules, but the prediction of the designed FES has become more accurate. 

\section{Summary and Conclusions}\label{con}
We have developed a fuzzy rule based expert system to predict the chances of having prostate cancer (PC) from a given set of input parameters. For this FES, we used age, PSA, PV and $\%$FPSA as the inputs to calculate prostate cancer risk (PCR) as the output. Existing literature on FES for PC, has mainly dealt with first three of the above mentioned inputs. The cutoff of $\%$FPSA is also a very crucial parameter along the PSA value, as it can reduce avoidable biopsy tests significantly. Most important advantages, apart from the cost of biopsy, the patient who do not need biopsy will be free from uneasiness and worries of the procedure, as well as the possible aftermath medical complications \cite{labmed}. Therefore, inclusion of this parameter in our FES was essential and it has altered the outcome appreciably. For a case study of $119$ patients, our FES has correctly predicted for $82$ patients ($68.91\%$). It is to be noted that, true prediction percentage for positive biopsy cases is excellent ($73.77\%$) whereas for the negative biopsy cases, it deviates slightly. The salient feature of our FES is that, the true prediction for positive biopsy is much higher compared to negative biopsy cases. This may cause an unnecessary biopsy test (along the with the pain associated with it) but it will certainly save a life. One should keep in mind that the FES designed here is not to confirm whether a person is having PC or not, but to assist a doctor to take a decision whether to go for biopsy or not. Since, four inputs are considered for this FES, number of rules has increased which makes it little complicated to start with. Once the burden of deciding these rules based on experts' opinion are done, it does the prediction efficaciously. On the other hand, the behaviour of different bio-markers varies drastically from race to race, from demographic region to region, family history, life style, food habits and also depends on many more hidden variables. So, modification in cutoff values has to be done accordingly to use the FES. Hybridizing our FES with other AI techniques and incorporating more experts' opinions may lead to an improved result.
\section*{Acknowledgements}
Authors would like to thank Dr. Bidesh Karmakar and Dr. Aniruddha Dewri for the useful discussions made with them during the manuscript preparation.

\bibliography{biblio}
\clearpage
\begin{center}
\begin{longtable}{*{12}c}
\caption{Data set containing the information of age, PSA, PV and $\%$FPSA values and biopsy results for 119 patients given in the reference \cite{saritas}. PCR values are calculated by our FES using Matlab fuzzy logic toolbox.}
\label{tab:6}\\  \hline
Age  & PSA  & PV  & \multirow{2}{*}{\%FPSA} & Biopsy & \multirow{2}{*}{PCR (\%)} & Age & PSA & PV  &  \multirow{2}{*}{\%FPSA} & Biopsy &  \multirow{2}{*}{PCR (\%)}  \\ 
(year) & (ng/ml) & (ml) &  & result &  & (year) & (ng/ml) & (ml) &  & result & \\  \hline 
\endfirsthead 
\multicolumn{12}{c}%
{{\bfseries \tablename\ \thetable{} -- continued from previous page}} \\ \hline 
Age  & PSA  & PV  & \multirow{2}{*}{\%FPSA} & Biopsy &  \multirow{2}{*}{PCR (\%)} & Age & PSA & PV  &  \multirow{2}{*}{\%FPSA} & Biopsy &  \multirow{2}{*}{PCR (\%)}  \\ 
(year) & (ng/ml) & (ml) &  & result &  & (year) & (ng/ml) & (ml) &  & result & \\ \hline
\endhead 
\hline \multicolumn{12}{r}{{Continued on next page}} \\  \hline
\endfoot
\hline
\endlastfoot
44 & 7.6 & 38 & 10.53 & Negative & 45.42 & 67 & 15.93 & 69 & 6.09 & Positive & 74.82  \\ \vspace{.2cm}
51 & 6.76 & 15 & 4.14 & Positive & 57.78 & 67 & 28 & 47 & 15.00 & Positive & 74.15   \\ \vspace{.2cm}
51 & 44 & 83 & 31.82 & Positive & 30.00 & 68 & 5.09 & 47 & 2.36 & Negative & 42.96  \\ \vspace{.2cm}
53 & 4.5 & 39 & 18.89 & Negative & 19.96 & 68 & 5.51 & 45 & 11.25 & Negative & 21.28  \\% \vspace{.2cm}
53 & 5.83 & 25 & 6.86 & Negative & 53.32 & 68 & 7.2 & 33 & 3.61 & Positive & 67.21   \\\vspace{.2cm}
53 & 8.34 & 25 & 7.43 & Negative & 73.83 & 68 & 9.25 & 91 & 3.57 & Positive & 74.03   \\ %\vspace{.2cm}
54 & 5.62 & 28 & 14.95 & Negative & 21.96 & 68 & 12.1 & 61 & 16.12 & Negative & 74.25   \\ \vspace{.2cm}
54 & 17.3 & 90 & 27.46 & Negative & 30.00 & 68 & 23.7 & 109 & 10.04 & Positive & 73.55   \\% \vspace{.2cm}
54 & 17.3 & 45 & 8.90 & Positive & 73.91 & 68 & 140 & 117 & 14.29 & Positive & 74.80  \\ \vspace{.2cm}
55 & 10.51 & 54 & 22.45 & Negative & 23.57 & 68 & 140 & 54 & 3.29 & Positive & 74.70   \\ %\vspace{.2cm}
56 & 8.9 & 26 & 34.16 & Negative & 18.80 & 69 & 8.8 & 34 & 8.98 & Positive & 74.40   \\ \vspace{.2cm}
56 & 9.05 & 39 & 8.51 & Positive & 74.07 & 69 & 11.06 & 38 & 29.84 & Negative & 26.64   \\% \vspace{.2cm}
56 & 16 & 146 & 8.44 & Negative & 74.07 & 69 & 15.31 & 74 & 30.57 & Positive & 30.00   \\ \vspace{.2cm}
57 & 12.56 & 52 & 65.84 & Negative & 30.00 & 69 & 61 & 46 & 9.93 & Negative & 73.65   \\ %\vspace{.2cm}
58 & 4.48 & 67.5 & 16.07 & Negative & 16.09 & 69 & 70.56 & 45 & 6.02 & Positive & 73.98   \\ \vspace{.2cm}
58 & 4.62 & 48 & 11.04 & Negative & 17.62 & 69 & 146 & 29 & 7.33 & Positive & 75.10   \\ %\vspace{.2cm}
58 & 5.2 & 58 & 23.46 & Negative & 12.80 & 70 & 5.39 & 120 & 19.11 & Negative & 23.58   \\ \vspace{.2cm}
58 & 16.39 & 27 & 92.07 & Negative & 30.00 & 70 & 5.39 & 42 & 12.80 & Negative & 19.91   \\% \vspace{.2cm}
59 & 0.28 & 168 & 42.86 & Negative & 13.30 & 70 & 13 & 40 & 15.46 & Negative & 74.39   \\ \vspace{.2cm}
59 & 8.36 & 55 & 7.54 & Positive & 74.31 & 70 & 13.95 & 119 & 13.76 & Negative & 74.02   \\% \vspace{.2cm}
59 & 18.2 & 77 & 17.75 & Negative & 73.76 & 70 & 19.2 & 44 & 10.10 & Positive & 73.49   \\ \vspace{.2cm}
59 & 19.48 & 79 & 25.00 & Positive & 30.00 & 70 & 21.94 & 29 & 7.11 & Positive & 75.17   \\% \vspace{.2cm}
59 & 22.51 & 42 & 7.02 & Negative & 74.22 & 70 & 27.7 & 63 & 8.99 & Negative & 74.40   \\ \vspace{.2cm}
59 & 22.65 & 66 & 10.82 & Negative & 73.88 & 71 & 6.08 & 48 & 21.38 & Positive & 13.52   \\% \vspace{.2cm}
60 & 6.58 & 65 & 14.74 & Negative & 24.82 & 71 & 12.64 & 50 & 7.99 & Positive & 74.39   \\\vspace{.2cm}
60 & 10.6 & 30 & 16.79 & Positive & 61.27 & 71 & 22 & 57 & 12.00 & Positive & 74.59   \\ %\vspace{.2cm}
60 & 11.45 & 46 & 19.48 & Negative & 53.47 & 72 & 6.64 & 32 & 27.41 & Negative & 12.43   \\ \vspace{.2cm}
60 & 14.79 & 38 & 6.90 & Positive & 74.39 & 72 & 13.31 & 33 & 3.83 & Positive & 74.40   \\ %\vspace{.2cm}
60 & 15.51 & 35 & 21.02 & Negative & 30.00 & 72 & 13.31 & 33 & 3.76 & Positive & 74.40   \\ \vspace{.2cm}
61 & 4.6 & 37 & 10.87 & Negative & 25.02 & 72 & 20 & 48 & 7.90 & Positive & 74.22   \\ %\vspace{.2cm}
61 & 10.33 & 62 & 25.36 & Negative & 23.78 & 72 & 46 & 36 & 10.70 & Positive & 73.81   \\ \vspace{.2cm}
61 & 10.36 & 35 & 19.79 & Negative & 47.88 & 72 & 77 & 48 & 8.31 & Positive & 74.22   \\ \vspace{.2cm}
61 & 10.59 & 56 & 17.00 & Positive & 60.72 & 73 & 4.65 & 41 & 41.94 & Negative & 12.52   \\ \vspace{.2cm}
61 & 18.3 & 62 & 6.99 & Positive & 74.46 & 73 & 7.25 & 19 & 5.52 & Negative & 67.73   \\ \vspace{.2cm}
62 & 6.12 & 52 & 24.18 & Negative & 12.86 & 73 & 7.6 & 74 & 31.32 & Positive & 12.09   \\ \vspace{.2cm}
62 & 6.2 & 25 & 4.35 & Positive & 56.05 & 73 & 19 & 90 & 6.84 & Positive & 73.98   \\ \vspace{.2cm}
62 & 8.37 & 43 & 11.23 & Negative & 40.23 & 73 & 29.52 & 91 & 9.82 & Negative & 73.73   \\ \vspace{.2cm}
62 & 8.79 & 45 & 10.92 & Positive & 48.39 & 73 & 47.4 & 87 & 15.89 & Positive & 74.11   \\ \vspace{.2cm}
62 & 20 & 53 & 5.20 & Positive & 74.55 & 74 & 12.52 & 27 & 11.82 & Negative & 74.47   \\ \vspace{.2cm}
62 & 51.74 & 29 & 6.80 & Positive & 74.55 & 74 & 150 & 54 & 16.67 & Positive & 74.09   \\ \vspace{.2cm}
63 & 8.8 & 31 & 22.50 & Positive & 19.22 & 75 & 4.61 & 16 & 17.57 & Positive & 18.39   \\ \vspace{.2cm}
64 & 5.7 & 36 & 29.82 & Negative & 12.71 & 75 & 10 & 34 & 7.60 & Positive & 73.98   \\ \vspace{.2cm}
64 & 6.96 & 45 & 9.20 & Negative & 60.28 & 76 & 9.81 & 56 & 37.41 & Negative & 21.46   \\ \vspace{.2cm}
64 & 8 & 40 & 7.50 & Positive & 74.39 & 76 & 13.61 & 61 & 19.91 & Positive & 52.54   \\ \vspace{.2cm}
64 & 11.08 & 26 & 10.11 & Negative & 59.05 & 76 & 13.83 & 54 & 19.96 & Positive & 51.76   \\ \vspace{.2cm}
64 & 16.28 & 21 & 6.94 & Positive & 74.70 & 76 & 21 & 86 & 5.43 & Positive & 74.15   \\ \vspace{.2cm}
65 & 4.39 & 30 & 21.64 & Negative & 13.42 & 77 & 10 & 60 & 6.00 & Positive & 73.98   \\ \vspace{.2cm}
65 & 5.15 & 47 & 15.73 & Negative & 19.46 & 77 & 12.05 & 28 & 27.05 & Positive & 30.00   \\ \vspace{.2cm}
65 & 7.61 & 23 & 5.78 & Positive & 70.95 & 77 & 56 & 51 & 7.34 & Positive & 74.46   \\ \vspace{.2cm}
65 & 7.82 & 75 & 22.76 & Negative & 13.04 & 78 & 4.5 & 180 & 20.44 & Negative & 18.05   \\ \vspace{.2cm}
65 & 8.33 & 32 & 14.53 & Positive & 38.08 & 78 & 26.1 & 46 & 8.62 & Negative & 74.07   \\ \vspace{.2cm}
66 & 4.38 & 33 & 23.52 & Negative & 12.78 & 78 & 26.13 & 235 & 8.27 & Negative & 74.96   \\ \vspace{.2cm}
66 & 6.72 & 61 & 13.84 & Positive & 25.38 & 78 & 31.6 & 57 & 8.86 & Negative & 74.50   \\ \vspace{.2cm}
66 & 7.65 & 89 & 23.66 & Negative & 12.90 & 79 & 17.1 & 41 & 7.60 & Negative & 74.31   \\ \vspace{.2cm}
66 & 9 & 74 & 18.89 & Positive & 53.84 & 80 & 69.51 & 28 & 28.77 & Positive & 30.00   \\ \vspace{.2cm}
66 & 9.86 & 49 & 23.83 & Negative & 21.69 & 81 & 4.5 & 28 & 21.56 & Positive & 13.45   \\ \vspace{.2cm}
67 & 4.39 & 28 & 0.91 & Negative & 23.99 & 81 & 68.36 & 52 & 35.27 & Positive & 30.00   \\ \vspace{.2cm}
67 & 5.65 & 24 & 10.27 & Positive & 46.65 & 88 & 10.4 & 32 & 7.50 & Positive & 74.22   \\ \vspace{.2cm}
67 & 6.24 & 65 & 21.96 & Negative & 13.31 &  &  &  &  &  &    \\ \vspace{.2cm}
67 & 8.2 & 36 & 20.37 & Positive & 31.78 &  &  &  &  &  &    \\ \vspace{.2cm}
67 & 9.68 & 41 & 7.44 & Positive & 74.18 &  &  &  &  &  &    \\
\end{longtable}
\end{center}

\end{document}